%% file: main.tex
\newcommand{\CLASSINPUTbottomtextmargin}{4.16cm}

\def\review{1} 

\documentclass[conference,a4paper]{IEEEtran}
\usepackage[left=1.43cm,right=1.43cm,top=1.8cm,bottom=4.16cm]{geometry}
\IEEEoverridecommandlockouts
\usepackage{cite}
\usepackage{amsmath,amssymb,amsfonts,bbm}
\usepackage{graphicx}
\usepackage{textcomp}
\usepackage[dvipsnames]{xcolor}
\usepackage{multirow}
\usepackage{subcaption}
\usepackage{caption}\usepackage{tikz}
\usepackage{tkz-tab}
\usetikzlibrary{automata,arrows,positioning,calc}
\usetikzlibrary{shapes,snakes}
\usetikzlibrary{arrows}
\usepackage{dsfont}
\usepackage{soul}
\usepackage{subfiles}
\usepackage{comment}
\usepackage{algorithm}
\usepackage{algpseudocode}
\usepackage{amsmath}
\usepackage{amssymb}
\usepackage{amsthm}
\usepackage{booktabs}
\usepackage{epstopdf} 
\usepackage{array}
\usepackage{diagbox}
\usepackage{eso-pic}

\def\BibTeX{{\rm B\kern-.05em{\sc i\kern-.025em b}\kern-.08em
    T\kern-.1667em\lower.7ex\hbox{E}\kern-.125emX}}

\if\review1
\usepackage{todonotes}
\else
\usepackage[disable]{todonotes}
\fi

\usepackage{tikz}
\usetikzlibrary{fit,calc}

\colorlet{pink}{red!40}
\colorlet{blue}{cyan!50}

\usepackage[acronym]{glossaries}
\makeglossaries
\newacronym{dvm}{DVM}{Drain Voltage Modulation}
\newacronym{dv}{DV}{Drain Voltage}
\newacronym{ran}{RAN}{Radio Access Network}
\newacronym{5g}{5G}{Fifth Generation}
\newacronym{6g}{6G}{Sixth Generation}
\newacronym{bs}{BS}{Base Station}
\newacronym{pa}{PA}{Power Amplifier}
\newacronym{rf}{RF}{Radio Frequency}
\newacronym{dc}{DC}{Direct Current}
\newacronym{apt}{APT}{Average Power Tracking}
\newacronym{et}{ET}{Envelope Tracking}
\newacronym{eer}{EER}{Envelope Elimination and Restoration}
\newacronym{papr}{PAPR}{Peak-to-Average Power Ratio}
\newacronym{tdd}{TDD}{Time-Division Duplex}
\newacronym{ofdm}{OFDM}{Orthogonal Frequency-Division Multiplexing}
\newacronym{nr}{NR}{New Radio}
\newacronym{ru}{RU}{Radio Unit}
\newacronym{du}{DU}{Distributed Unit}
\newacronym{mac}{MAC}{Media Access Control}
\newacronym{tddvm}{TDD-DVM}{TDD-Aligned Drain Voltage Modulation}
\newacronym{isd}{ISD}{Inter-Site Distance}
\newacronym{ue}{UE}{User Equipment}
\newacronym{prb}{PRB}{Physical Resource Block}
\newacronym{irc}{IRC}{Interference Rejection Combining}
\newacronym{rbg}{RBG}{Resource Block Group}
\newacronym{sinr}{SINR}{Signal-to-Interference-plus-Noise Ratio}
\newacronym{miesm}{MIESM}{Mutual Information Equivalent SINR Mapping}
\newacronym{la}{LA}{Link Adaptation}
\newacronym{mcs}{MCS}{Modulation and Coding Scheme}
\newacronym{harq}{HARQ}{Hybrid Automatic Repeat Request}
\newacronym{bler}{BLER}{Block Error Rate}
\newacronym{bla}{BLA}{Baseline Link Adaptation}
\newacronym{polite}{POLITE}{Power Optimization for Low Interference and Throughput Enhancement}
\newacronym{dl}{DL}{Downlink}
\newacronym{ul}{UL}{Uplink}
\newacronym{psdvm}{PS-DVM}{Per-Slot Drain Voltage Modulation}
\newacronym{bo}{BO}{Bayesian Optimization}
\newacronym{bogp}{BOGP}{Bayesian Optimization with Gaussian Processes}
\newacronym{rrm}{RRM}{Radio Resource Management}

\begin{document}
\AddToShipoutPictureBG*{%
  \AtPageLowerLeft{%
    \put(\LenToUnit{0.5\paperwidth},\LenToUnit{1cm}){%
      \makebox[0pt]{\parbox{0.85\paperwidth}{\centering\footnotesize
      This work has been submitted to the IEEE for possible publication. Copyright may be transferred without notice, after which this version may no longer be accessible.}}}}}
      
\bstctlcite{IEEEexample:BSTcontrol}

\title{Reducing 6G Energy Consumption: TDD-Aligned Drain Voltage Modulation for Power Amplifiers}

\author{Meizhen Zhuo$^{\star\dagger}$, Gunter Wolff$^{\S}$, Stephan ten Brink$^{\dagger}$, Silvio Mandelli$^{\star}$ \\
\IEEEauthorblockA{$^{\star}$\emph{Nokia Bell Labs, Stuttgart, Germany}}
\IEEEauthorblockA{$^{\S}$\emph{Nokia Technology Center, Ulm, Germany}}
\IEEEauthorblockA{$^{\dagger}$\emph{Institute of Telecommunications, University of Stuttgart, Stuttgart, Germany}}

\IEEEauthorblockA{E-mail: meizhen.zhuo@nokia.com}
}

\maketitle

\thispagestyle{plain}
\pagestyle{plain}

\subfile{sections/0_abstract}


\subfile{sections/1_intro}

\subfile{sections/2_system_model}

\subfile{sections/3_proposed_urdvm}

\subfile{sections/4_simulations}

\subfile{sections/5_conclusions}

\bibliographystyle{IEEEtran}
\bibliography{references}

\end{document}

%% file: sections/0_abstract.tex
\begin{abstract}

The \gls{pa} of a base station operates inefficiently at partial load, where back-off from the peak output power wastes supply headroom. Adapting the \gls{pa} drain voltage to the instantaneous signal recovers this loss but requires frequent switching, increasing hardware cost. We propose \gls{tddvm}, a drain voltage modulation scheme that instead selects the drain voltage level once per \gls{tdd} period based on the downlink resource load and switches the voltage during the uplink window. This is enabled by the authors' previous work on POLITE link adaptation, which spreads the transmit power over resources left unused at partial load and moves the \gls{pa} into a regime where a lower drain voltage suffices. In a 3GPP Urban Macro system-level simulation across traffic loads, we compare \gls{tddvm} against baseline approaches under conventional and POLITE link adaptation. At equal delivered throughput, \gls{tddvm} with POLITE keeps the packet failure ratio of the full drain voltage while consuming up to $35\%$ less power than it under conventional link adaptation and up to $7\%$ less than it under POLITE, capturing most of the gain of an idealized reference at moderate loads and roughly half at the highest load.

\end{abstract}

\glsreset{pa}
\glsreset{bs}
\glsreset{tddvm}
\glsreset{tdd}
\glsreset{dl}
\glsreset{ul}
\glsreset{polite}
\glsreset{psdvm}
\glsreset{bla}
\glsreset{la}

%% file: sections/1_intro.tex
\section{Introduction}
\label{sec:introduction}

Improving the energy efficiency of \glspl{ran} has emerged as a central challenge for \gls{5g}-Advanced and future \gls{6g} systems. \glspl{bs} dominate energy consumption in cellular networks, accounting for over 70\% of total network power consumption\cite{piovesan2022machine}. Within a \gls{bs}, the \gls{pa} is responsible for a substantial share of the power dissipation\cite{lopez2022survey}, making it a primary target for energy optimization.

A fundamental source of energy inefficiency arises from the mismatch between network design requirements and real-world operating conditions. Mobile networks are dimensioned to support peak traffic demand, whereas actual traffic load varies significantly over time and across locations\cite{ozturk2021energy}. Consequently, \glspl{bs} are typically over-provisioned and operate well below peak capacity for much of the time.

At the \gls{pa} level, this under-utilization keeps the \gls{pa} backed off from its peak output power for extended periods. This regime exposes a key limitation of fixed-supply \gls{pa} operation. Under a fixed drain voltage, \gls{pa} efficiency degrades substantially at output-power back-off because the supply headroom reserved for peak-power operation leads to excess \gls{dc} power dissipation\cite{asbeck2016comes}. Lowering the drain voltage at reduced output power can improve \gls{pa} efficiency\cite{asbeck2016comes}, as illustrated by the measured \gls{pa} efficiency in Fig.~\ref{fig:pa_eff}, while high output power still requires sufficient supply headroom to maintain linearity and output capability.

This has motivated \gls{pa} supply modulation techniques, which dynamically adapt the supply voltage to the operating power requirement, improving back-off efficiency while retaining peak power capability\cite{nogales2023800}. These techniques differ mainly in how fast the supply voltage is updated. \gls{et} follows the instantaneous signal envelope, recovering efficiency under the high \gls{papr} of \gls{ofdm}-based waveforms at the cost of a fast, wideband supply path and linearization to contain tracking-induced distortion\cite{asbeck2016comes}. \gls{apt} instead follows the average transmit power over a tracking interval\cite{park2015analysis}, which adapts more coarsely but requires only a slow supply path. In both cases, the supply is modulated at the \gls{rf} transmitter level and driven by the transmitted signal itself.

A complementary line of work reduces \gls{bs} energy by exploiting traffic variation at the network level. In the time and frequency domains, widely studied approaches include \gls{bs} sleep modes\cite{yang2017hysteretic}, symbol- or slot-level power shutdown\cite{tan2022energy}, and carrier shutdown in \gls{5g} \gls{nr}\cite{kundu2025toward}. In the power domain, energy saving is addressed through transmit-power adaptation, as in \gls{polite}\cite{mandelli2021power, mandelli2021powerdelay}, which spreads a transmission over more resources at lower power whenever the cell is not fully loaded.

Adapting the drain voltage to the instantaneous signal requires expensive hardware, since the supply must follow the signal almost instantaneously. We instead update it on a slow time scale, once per \gls{tdd} period, and place the update in the \gls{ul} period, which carries no \gls{dl} transmission and allows a slow ramp without signal distortion, as illustrated in Fig.~\ref{fig:urdvm_timing}. The supply thus has the whole \gls{ul} period to settle, orders of magnitude longer than the signal-driven timescales of \gls{et} and \gls{apt}, so a far slower and cheaper supply path suffices. We address the problem of minimizing the time-averaged consumed \gls{pa} power by selecting the drain voltage level once per \gls{tdd} period, while still serving the offered \gls{dl} traffic. In this paper, we propose the load-based \gls{tddvm} algorithm, which selects the \gls{pa} drain voltage based on the current resource load before the next \gls{dl} transmission opportunity.
Our main contributions are:
\begin{itemize}
    \item We align the drain voltage adaptation with the \gls{tdd} frame, ramping up/down the \gls{pa} supply once per \gls{tdd} period during the \gls{ul}, lowering the hardware cost of drain voltage switching.
    \item We incorporate \gls{polite} \gls{la} into \gls{tddvm}. By lowering the transmit power whenever the cell is not fully loaded, \gls{polite} moves the \gls{pa} into a region where a lower drain voltage suffices, which \gls{tddvm} exploits to reduce the consumed power.
    \item We evaluate our proposal against other baselines with extensive system-level simulations, analyzing consumed power, delivered throughput, and packet failure ratio.
\end{itemize}

%% file: sections/2_system_model.tex
\section{System Model}
\label{sec:system_model}

We consider a multi-cell cellular network operating in \gls{tdd} mode, in which \gls{la} determines the per-resource transmit power, which in turn drives the drain-voltage-dependent \gls{pa} power consumption.

\subsection{TDD Frame Structure}
\label{subsec:tdd_structure}

Transmission follows a \gls{tdd} frame structure that repeats periodically, multiplexing \gls{ul} and \gls{dl} in the time domain over a shared carrier. Each \gls{tdd} period comprises $N_{\mathrm{TDD}}$ slots of duration $T_s$, of which $N_{\mathrm{UL}}$ are assigned to the \gls{ul} and the remaining
\begin{equation}
    N_{\mathrm{DL}} = N_{\mathrm{TDD}} - N_{\mathrm{UL}}
    \label{eq:ndl}
\end{equation}
to the \gls{dl}. The pattern is semi-static and common to all cells, so that the \gls{ul} and \gls{dl} periods are aligned across the network. 

\subsection{Traffic, Resource, and Link Adaptation Model}
\label{subsec:la_resource}

We describe the model for a single cell, which serves a set of \glspl{ue} $\mathcal{U}$, each requesting downloads to be delivered within a deadline $\tau$. We denote by $\mathcal{R}$ the set of resources available for scheduling in one slot, so that the \gls{dl} resources within one \gls{tdd} period are the $|\mathcal{R}|$ resources in each of the $N_{\mathrm{DL}}$ \gls{dl} slots.

Let $P^{r}(t)$ denote the transmit power on resource $r \in \mathcal{R}$ in slot $t$. For a \gls{ue} $u \in \mathcal{U}$ scheduled on resource $r$, the per-resource \gls{sinr} $\gamma_u^{r}(t)$ is determined by $P^{r}(t)$, the channel gains, the inter-cell interference, and noise. The values $\gamma_u^{r}(t)$ within a transmission are mapped to a single effective \gls{sinr}, e.g., through \gls{miesm}~\cite{brueninghaus2005link}.

Based on the effective \gls{sinr}, the \gls{la} unit selects the \gls{mcs} so that the \gls{bler} does not exceed a target $\mathrm{BLER}_{\mathrm{tgt}}$. A per-resource proportional-fair scheduler distributes the resources across the active \glspl{ue}. To support the load estimation used below, we compute the smoothed achievable spectral efficiency of \gls{ue} $u$ as
\begin{equation}
s_u(t) = \delta\, s_u(t-1) + (1-\delta)\,\hat{s}_u(t),
\label{eq:se}
\end{equation}
where $\hat{s}_u(t)$ is the instantaneous achievable spectral efficiency reported by the \gls{la} unit and $\delta \in (0,1)$ is a smoothing factor.

We measure the delivered traffic against its deadline. The average delivered throughput of \gls{ue} $u$ is
\begin{equation}
\Gamma_u = \frac{c_u}{T_{\mathrm{run}}},
\label{eq:goodput}
\end{equation}
where $c_u$ is the traffic correctly received by \gls{ue} $u$ within the deadline $\tau$ and $T_{\mathrm{run}}$ is the total duration of the run. The packet failure ratio is
\begin{equation}
\phi = 1 - \frac{\sum_{u\in\mathcal{U}} c_u}{\sum_{u\in\mathcal{U}} o_u},
\label{eq:phi}
\end{equation}
where $o_u$ is the traffic offered to \gls{ue} $u$, so that $\phi$ is the fraction of offered traffic of the cell not delivered within the deadline.

This work considers the following two \gls{la} schemes.

\subsubsection{\gls{bla}}
\label{subsec:bla}

The baseline scheme maximizes the spectral efficiency of each scheduled \gls{ue} by selecting the highest-rate \gls{mcs} whose \gls{bler} does not exceed the target $\mathrm{BLER}_{\mathrm{tgt}}$, and transmits at the maximum available power. This minimizes the number of occupied resources per transmission. Since the transmit power is kept at its maximum regardless of the traffic load, the \gls{pa} output power remains high, so that the highest drain voltage is required more often and more energy is consumed. \gls{bla} serves as one of the reference schemes in our evaluation.

\subsubsection{POLITE Link Adaptation}
\label{subsec:polite_la}

In contrast to \gls{bla}, \gls{polite}~\cite{mandelli2021power} spreads a given amount of data over more resources at a lower per-resource transmit power $P^{r}(t)$, exploiting the resources left unused at partial load while preserving reliability. We adopt its delay-aware variant~\cite{mandelli2021powerdelay}, which additionally exploits the traffic delay budget when deciding how much to spread. \gls{polite} quantifies the resource utilization under \gls{bla}, defined as the ratio between the resources required by \gls{bla} to serve the traffic and those available in a slot, which reads
\begin{equation}
\beta(t) = \max\!\Big[\beta_{\min},\ \min\!\Big( \sum_{u\in\mathcal{U}}
\frac{\rho'_u(t)}{s_u(t)\,|\mathcal{R}|},\ 1 \Big)\Big],
\label{eq:beta}
\end{equation}

where $\rho'_u(t)$ is the required service rate of \gls{ue} $u$, computed in a delay-aware manner as the larger of its smoothed arrival rate and the throughput needed to clear its queued packets within their delay budgets~\cite{mandelli2021powerdelay}, and $\beta_{\min}$ is a lower bound that limits the power reduction.

When $\beta(t)<1$, \gls{polite} downgrades the \gls{mcs} of each \gls{ue} to the lowest rate in the ordered \gls{mcs} set that still reaches a fraction $\beta(t)$ of the rate that \gls{bla} would select, and then reduces the transmit power $P^{r}(t)$ to the minimum at which the more robust \gls{mcs} still meets the $\mathrm{BLER}_{\mathrm{tgt}}$. By transmitting at reduced power on more resources, \gls{polite} lowers the \gls{pa} output power at partial load, so that a lower drain voltage suffices more often. This opens the opportunity for the energy efficiency gain of drain voltage modulation, which the proposed \gls{tddvm} delivers.


\subsection{PA Power Consumption Model}
\label{subsec:pa_model}

The \gls{pa} can be operated at one of $K$ discrete drain voltage levels, indexed by $k\in\{1,\dots,K\}$ in increasing order of voltage, where level $K$  is the full drain voltage of the \gls{pa}. A higher level extends the linear range of the \gls{pa} and supports a larger maximum \gls{rf} output power, so that $p_{\max}(1)<p_{\max}(2)<\cdots<p_{\max}(K)$, and level $k$ can serve an \gls{rf} output power $p$ only if $p\le p_{\max}(k)$.

For each level $k$, let $\eta_k:\mathbb{R}\to[0,1]$ be the \gls{pa} efficiency, so that $\eta_k(p)$ is the efficiency at \gls{rf} output power $p$. The consumed \gls{dc} power at \gls{rf} output power $p$ is then
\begin{equation}
P_{\mathrm{DC}}(p, k) =
\begin{cases}
P_{\mathrm{idle}}, & p = 0, \\[3pt]
\dfrac{p}{\eta_k(p)}, & 0 < p \le p_{\max}(k),
\end{cases}
\label{eq:pdc}
\end{equation}
where $P_{\mathrm{idle}}$ is the power drawn at zero \gls{rf} output. The dependency of $\eta_k$ on the output power and drain voltage level is the basis of the energy optimization developed in this work.

The transmit power is balanced across the $N_{\mathrm{tx}}$ transmit antennas, so the \gls{rf} output power of antenna $a$ in slot $t$ is
\begin{equation}
    p_a(t)=\frac{1}{N_{\mathrm{tx}}}\sum_{r\in \mathcal{R}(t)} P^{r}(t),
    \label{eq:pa}
\end{equation}
where $\mathcal{R}(t)\subseteq\mathcal{R}$ is the set of \gls{dl} resources scheduled in slot $t$. Because each antenna is driven by its own \gls{pa}, the total \gls{pa} power consumed by the cell is
\begin{equation}
    P_{\mathrm{cell}}(t)=\sum_{a=1}^{N_{\mathrm{tx}}}
    P_{\mathrm{DC}}\big(p_a(t),\,k(t)\big),
    \label{eq:pcell}
\end{equation}
where a single drain voltage level $k(t)$ is applied across all antennas of the cell.

\subsection{Problem Formulation}
\label{subsec:problem}

Building on the model above, our objective is to minimize the long-term average \gls{dl} \gls{pa} power while still serving the offered \gls{dl} traffic. For each cell we select the drain voltage levels $\{k(t)\}$ that solve
\begin{equation}
\begin{aligned}
\min_{\{k(t)\}}\quad & \frac{1}{T}\sum_{t=1}^{T} P_{\mathrm{cell}}(t)\\
\text{s.t.}\quad
& p_a(t)\le p_{\max}\!\bigl(k(t)\bigr),\quad \forall a,\,t,\\
& k(t)\ \text{constant within each \gls{tdd} period},
\end{aligned}
\label{eq:problem}
\end{equation}
where the sum runs over the $T$ \gls{dl} slots of the run, $P_{\mathrm{cell}}(t)$ is given by \eqref{eq:pcell}, and the constraint guarantees that the selected drain voltage supports the power requested by the scheduler. Every feasible selection therefore serves the offered traffic exactly as the full drain voltage does, and permanently selecting a lower level is infeasible whenever the requested power exceeds its $p_{\max}$. A delivery constraint becomes necessary only when the level is selected from a load estimate, whose prediction errors can break this guarantee, and is introduced in \eqref{eq:bo}.

%% file: sections/3_proposed_urdvm.tex
\section{Drain Voltage Selection Schemes}
\label{sec:ur-dvm}

This section describes how the drain voltage level $k(t)$ is selected over time. We first present several reference schemes and then the proposed \gls{tddvm}. 

\subsection{Reference Schemes}
\label{subsec:ref_schemes}

\subsubsection{Fixed Drain Voltage}
A fixed-voltage reference keeps the drain voltage at a constant level $k$ over time. Since level $k$ can support an output power only up to $p_{\max}(k)$, the scheduler can allocate resources in a slot only as long as the resulting output power $p_a(t)$ stays within this limit, and the traffic that does not fit is postponed. The consumed power then follows from \eqref{eq:pdc} evaluated at the fixed level.

\subsubsection{Per-Slot DVM (PS-DVM)}
\gls{psdvm} selects the drain voltage level independently in every \gls{dl} slot. Among the levels supporting the output power $p_a(t)$ in that slot, it selects the one with the lowest consumed power. Since the level follows the instantaneous transmit power, \gls{psdvm} requires switching the drain voltage on the per-slot timescale, at additional hardware cost.

\subsubsection{Genie-Aided}
The Genie-Aided scheme selects one drain voltage level per \gls{tdd} period, assuming perfect knowledge of the transmissions in the upcoming \gls{dl} period. It takes the peak output power over that period and, following the same rule as \gls{psdvm}, selects the most energy-efficient level able to support it. Constrained to a single level per period, it cannot outperform the per-slot \gls{psdvm}, but it avoids the per-slot voltage switching cost. Since no realizable scheme can know this peak in advance, it is not implementable and serves as a bound for per-period selection, which the proposed \gls{tddvm} (Section~\ref{subsec:tdddvm}) aims to approach.

\subsection{Proposed Load-Based TDD-DVM With POLITE}
\label{subsec:tdddvm}

\begin{figure}[h]
  \vspace{-0pt}
  \centering
  \includegraphics[width=\columnwidth]{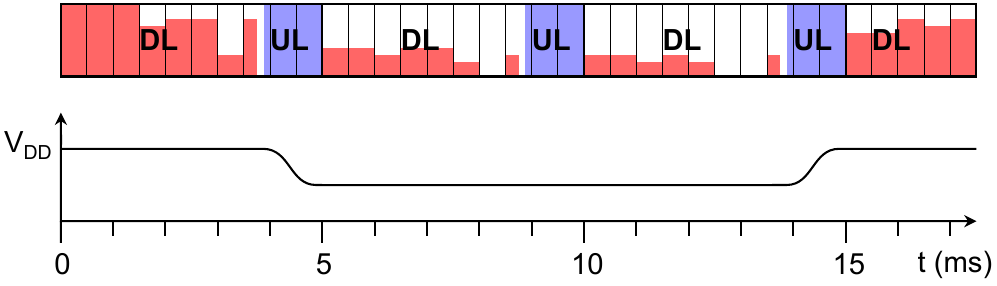}
  \caption{Timing of \gls{tddvm} for $K=2$ levels. The top strip illustrates the traffic load across slots. The drain voltage $V_{\mathrm{DD}}$ is ramped during each \gls{ul} period when the selected level changes, and held constant over the following \gls{dl} period.}
  \label{fig:urdvm_timing}
\end{figure}

Our proposed \gls{tddvm} updates the drain voltage once per \gls{tdd} period. The drain voltage level is ramped up or down during the \gls{ul} period and kept fixed throughout the following \gls{dl} period, so that the transition is completed before \gls{dl} transmission begins, as illustrated in Fig.~\ref{fig:urdvm_timing}. Since no \gls{dl} signal is present during the ramp, it introduces no distortion, while the per-period adaptation avoids the hardware cost of frequent switching. In return, the scheme is not autonomous and relies on signaling from the \gls{du} to align the ramping with the \gls{tdd} periods.

The \gls{du} decides the target level from a resource load $\Lambda(t)$ that reflects how heavily the cell is loaded,
\begin{equation}
\Lambda(t) = \frac{R_{\mathrm{req}}(t)}{N_{\mathrm{av}}},
\qquad
R_{\mathrm{req}}(t) = \sum_{u\in\mathcal{U}} \frac{Q_u(t)}{s_u(t)},
\label{eq:lambda}
\end{equation}
where $Q_u(t)$ is the traffic buffered for \gls{ue} $u$ at the start of the \gls{tdd} period and $s_u(t)$ its smoothed achievable spectral efficiency from \eqref{eq:se}, so that $R_{\mathrm{req}}(t)$ is the number of \gls{dl} resources needed to clear the current backlog of all \glspl{ue} in the cell, and $N_{\mathrm{av}}=|\mathcal{R}|\cdot N_{\mathrm{DL}}$ is the number of \gls{dl} resources available over one \gls{tdd} period.

A hysteresis rule with thresholds $\theta_{\mathrm{low}}<\theta_{\mathrm{high}}$ maps this resource load to the level decision. When $\Lambda(t)$ falls below $\theta_{\mathrm{low}}$, the resource load is light and the level is decreased by one step to save power, unless it is already at the lowest. When $\Lambda(t)$ exceeds $\theta_{\mathrm{high}}$, the resource load is heavy and the level is increased by one step to preserve the output-power capability, unless it is already at the highest. Otherwise the level is left unchanged, so that small resource load fluctuations do not trigger switching. The thresholds are tuned by \gls{bo}, as described in Section~\ref{subsec:bayesianoptimization}.

Algorithm~\ref{alg:tdddvm} summarizes the complete scheme, with the per-period drain voltage update of \gls{tddvm} as the outer loop and \gls{polite} \gls{la} per slot as the inner loop, setting the \gls{mcs} and per-resource transmit power. The per-slot consumed power $P_{\mathrm{cell}}(t)$ is averaged over the simulation to give the time-averaged consumed power reported in Section~\ref{sec:results}.

\begin{algorithm}[t]
\caption{TDD-DVM with POLITE Link Adaptation}
\label{alg:tdddvm}
\begin{algorithmic}[1]
\State initialize drain voltage level $k$
\For{each \gls{tdd} period}
    \State \textit{// UL period: TDD-DVM ramps the drain voltage}
    \State compute $\Lambda(t)$ from \eqref{eq:lambda} at the start of the period
    \If{$\Lambda(t) < \theta_{\mathrm{low}}$}
        \State $k \gets \max(k-1,\,1)$
    \ElsIf{$\Lambda(t) > \theta_{\mathrm{high}}$}
        \State $k \gets \min(k+1,\,K)$
    \EndIf
    \State \textit{// DL period: POLITE runs per slot with level $k$ fixed}
    \For{each \gls{dl} slot $t$ in the period}
        \State compute $\beta(t)$ from \eqref{eq:beta}
        \State schedule resources by proportional fairness
        \State set each scheduled \gls{ue}'s \gls{mcs} and transmit power
               $P^{r}(t)$ according to $\beta(t)$
        \State compute $p_a(t)$ from \eqref{eq:pa} and
               $P_{\mathrm{cell}}(t)$ from \eqref{eq:pcell}
    \EndFor
\EndFor
\end{algorithmic}
\end{algorithm}

\subsection{Bayesian Optimization (BO)}
\label{subsec:bayesianoptimization}

The two thresholds $\theta_{\mathrm{low}}$ and $\theta_{\mathrm{high}}$ control the trade-off between energy saving and the risk of operating at a drain voltage level too low to serve the traffic. We tune them for different traffic scenarios using \gls{bogp}, a sample-efficient black-box optimization method well suited to \gls{rrm} problems such as power control, spectrum allocation, and scheduling under dynamic and uncertain network conditions~\cite{maggi2021bayesian}. For a given traffic scenario, it iteratively selects threshold pairs to evaluate, balancing exploration of uncertain regions and exploitation of promising configurations.

Minimizing the consumed power alone would favor the lowest drain voltage regardless of whether the traffic is served, so the thresholds are tuned to minimize the consumed power subject to a delivery constraint,
\begin{equation}
\begin{aligned}
\min_{(\theta_{\mathrm{low}},\, \theta_{\mathrm{high}})} \quad
& \bar{P} = \frac{1}{T}\sum_{t=1}^{T} P_{\mathrm{cell}}(t) \\
\text{s.t.} \quad & \phi \le \phi_{\mathrm{ref}},
\end{aligned}
\label{eq:bo}
\end{equation}
where $\phi$ is the packet failure ratio of \eqref{eq:phi} and $\phi_{\mathrm{ref}}$ is the same ratio obtained by the full drain voltage at the same traffic load. The constraint keeps the delivery no worse than the full drain voltage would provide, so that the saving comes from lowering the voltage only where the traffic is still served. Repeating this over a range of traffic scenarios, separately for each \gls{la} strategy, yields the optimal threshold pair as a function of the traffic, which we fit with a regression curve. The thresholds are then selected from this curve according to the prevailing traffic, which the \gls{bs} can estimate by averaging the arriving bit rate over a sliding window, since all \gls{dl} traffic passes through its buffer.

%% file: sections/4_simulations.tex
\section{Numerical Evaluation}
\label{sec:results}

\subsection{Evaluation Setup}
\label{subsec:setup}

We evaluate the proposed scheme with a system-level simulator following the 3GPP 3D Urban Macro model. The network consists of a set of sites arranged in a hexagonal multi-site layout with an \gls{isd} of $d$. Each site comprises three $120^\circ$ sectors, and each sector is treated as an independent cell served by its own directional transceiver. A wrap-around topology is used to eliminate border effects and to provide a statistically homogeneous interference environment for all cells, and all results are averaged over the cells of the network. Each cell is equipped with $N_{\mathrm{tx}}$ co-located transmit antennas and operates in closed-loop spatial multiplexing mode. A set of \glspl{ue} $\mathcal{U}$ is dropped over the playground, each carrying $N_{\mathrm{rx}}$ receive antennas and an \gls{irc} receiver~\cite{3gpp36866}. The system operates at a carrier frequency $f_c$ over a bandwidth $B$ and follows the \gls{nr} numerology, and each cell serves FTP3 traffic and schedules the active \glspl{ue} with a proportional fair metric. The main simulation parameters are listed in Table~\ref{tab:params}.

\begin{table}[h]
\centering
\caption{Simulation Parameters}
\label{tab:params}
\begin{tabular}{ll}
\hline
Deployment & 3GPP 3D Urban Macro \\
\hline
Number of $120^\circ$ Cells  & 7 Sites $\times$ 3 Sectors = 21 \\
Inter-Site Distance $d$ & 500\,m \\
Carrier Frequency $f_c$, Bandwidth $B$  & 3.5\,GHz, 20\,MHz \\
Subcarrier Spacing & 30\,kHz \\
UL/DL Slots per Period $N_{\mathrm{UL}}$, $N_{\mathrm{DL}}$ & $2$, $8$ \\
Slot Duration $T_s$ & $0.5$\,ms \\
TX/RX Antennas $N_{\mathrm{tx}}$, $N_{\mathrm{rx}}$  & 8, 2 \\
Max Transmit Power & 49\,dBm \\
Drain Voltage Levels & $K=2$, $\{80,\,100\}\%$ of full DV \\
MCS Range & up to 64-QAM \\
FTP3 Users per Cell & 5 \\
FTP3 Packet Size & 0.4\,Mbit \\
FTP3 Packet Timeout $\tau$ & 4\,s \\
User Mobility & 3\,km/h \\
BLER Target & 0.1\\ 
Smoothing Factor $\delta$ & 0.05 \\
POLITE Minimum Ratio $\beta_{\min}$ & $0.5$ \\
BOGP Iterations & 50 \\
\hline
\end{tabular}
\end{table}

\begin{figure}[!h]
  \vspace{-7pt}
  \centering
  \includegraphics[width=0.76\columnwidth]{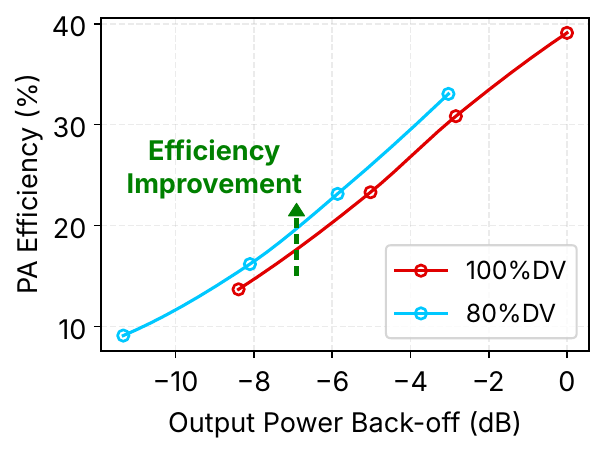}
  \caption{Measured \gls{pa} efficiency of both drain voltage levels versus output power back-off from the $100\%$DV peak.}
  \vspace{-6pt}
  \label{fig:pa_eff}
\end{figure}

We model the consumed \gls{dc} power of the \gls{pa} and additional transmitter components, referred to as \gls{pa} power in the following, through a lookup table over the \gls{rf} output power and drain voltage level, using $K=2$ discrete levels at $80\%$ and $100\%$ of the full \gls{dv}, denoted as $80\%$DV and $100\%$DV. Table~\ref{tab:pa} reports the measured characteristics of each level. For an output power $p>0$, the consumed power of \eqref{eq:pdc} is obtained as
\begin{equation}
P_{\mathrm{DC}}(p,k) = f_k(p) + P_{\mathrm{base}},
\label{eq:pdc_setup}
\end{equation}
where $f_k(p)$ is a spline interpolation of the measured points of level $k$ and $P_{\mathrm{base}}$ is the base power at nonzero output. The efficiency of \eqref{eq:pdc} then follows as $\eta_k(p)=p/\bigl(f_k(p)+P_{\mathrm{base}}\bigr)$, shown in
Fig.~\ref{fig:pa_eff} for the two levels. At zero output the \gls{pa} consumes $P_{\mathrm{idle}}$, as in \eqref{eq:pdc}.

\begin{table}[h]
\centering
\caption{Measured PA Characteristic}
\label{tab:pa}
\begin{tabular}{lrrrrr}
\hline
\multicolumn{6}{c}{$P_{\mathrm{base}} = 15.3\%$,\quad
$P_{\mathrm{idle}} = 8.9\%$\quad (both levels)} \\
\hline
\multicolumn{6}{c}{Level $k=1$ ($80\%$DV)} \\
\hline
RF power (\%) & $0.0$ & $7.4$ & $15.5$ & $25.9$ & $49.8$ \\
DC power (\%) & $13.9$ & $16.3$ & $22.1$ & $28.5$ & $43.6$ \\
\hline
\multicolumn{6}{c}{Level $k=2$ ($100\%$DV)} \\
\hline
RF power (\%) & $0.0$ & $14.5$ & $31.5$ & $52.0$ & $100.0$ \\
DC power (\%) & $17.3$ & $26.1$ & $37.5$ & $50.6$ & $84.7$ \\
\hline
\end{tabular}
\par\vspace{2pt}
\parbox{\columnwidth}{\footnotesize Note: RF power relative to the peak
output of $100\%$DV, DC values relative to its peak consumption. DC rows
report $f_k(\cdot)$, excluding $P_{\mathrm{base}}$. $P_{\mathrm{idle}}$ is the consumption of the inactive transmitter.}
\end{table}

\subsection{Energy Consumption}
\label{subsec:energy}

\begin{figure}[!t]
  \vspace{-6pt}
  \centering
  \includegraphics[width=\columnwidth]{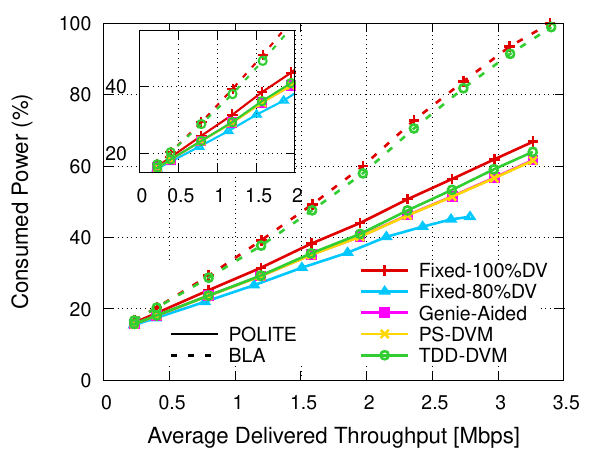}
  \caption{Consumed \gls{pa} power against the average delivered throughput per \gls{ue}. The inset magnifies the low-load region.}
  \label{fig:consumed}
\end{figure}

Fig.~\ref{fig:consumed} shows the consumed \gls{pa} power, normalized to the peak consumption of Fixed-$100\%$DV under \gls{bla}, against the average delivered throughput per \gls{ue}, swept over the offered traffic load. Neither fixed drain voltage resolves the trade-off between power consumption and capability. The Fixed-$100\%$DV wastes its supply headroom at every load, while the Fixed-$80\%$DV consumes the least but its curve terminates near $2.8$\,Mbps, where its power budget limits the resources the scheduler can allocate and part of the offered traffic is no longer delivered. Adapting the drain voltage resolves this trade-off. The proposed \gls{tddvm} with \gls{polite} lies between Fixed-$100\%$DV with \gls{polite} and the Genie-Aided bound, reducing the consumed power by up to $7\%$ at equal delivered throughput, against about $9\%$ for
\gls{psdvm} and the bound. \gls{tddvm} captures roughly half of this ideal gain at the highest load and most of it at moderate loads, at a fraction of the switching cost.

The comparison with \gls{bla} shows where this gain comes from. Under \gls{bla}, every transmission is served at maximum power, so the \gls{pa} output remains high whenever traffic is present and \gls{tddvm} consumes nearly as much as Fixed-$100\%$DV. \gls{polite} instead lowers the per-resource transmit power at partial load and moves the \gls{pa} into the region where $80\%$DV suffices, which \gls{tddvm} converts into the saving above. Relative to Fixed-$100\%$DV with \gls{bla}, the combination of \gls{polite} and \gls{tddvm} reduces the consumed power by up to $35\%$ at equal delivered throughput, of which the drain voltage adaptation contributes about $7\%$.

\subsection{Packet Failure Ratio}
\label{subsec:delivery}

\begin{figure}[!t]
  \vspace{-5pt}
  \centering
  \includegraphics[width=0.93\columnwidth]{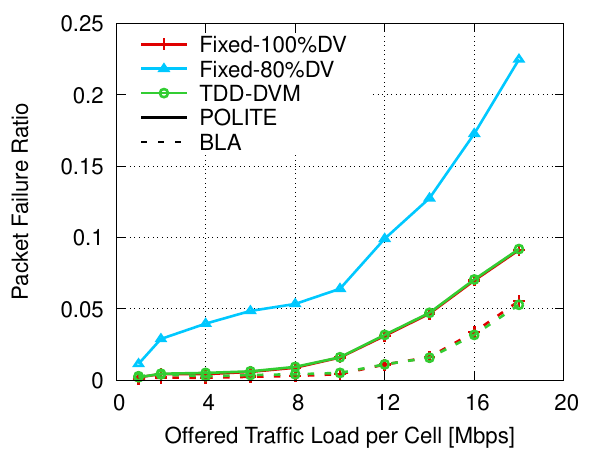}
  \caption{Packet failure ratio against the offered traffic load per cell. \gls{psdvm} and the Genie-Aided bound coincide with Fixed-$100\%$DV and are omitted.}
  \label{fig:failure}
\end{figure}

Fig.~\ref{fig:failure} shows the packet failure ratio of \eqref{eq:phi} against the offered traffic load per cell, and confirms that the energy saving of Fig.~\ref{fig:consumed} does not come at the expense of delivery. Under \gls{polite}, the failure ratio of the proposed \gls{tddvm} is indistinguishable from that of Fixed-$100\%$DV over the whole load range, so the delivery constraint of \eqref{eq:bo} is met. Fixed-$80\%$DV instead lies above at every load and reaches $0.22$ at the highest load, since its power budget limits the resources the scheduler can allocate and the postponed traffic exceeds its deadline. Lowering the voltage at all traffic loads is therefore not viable. The same behavior holds under \gls{bla}, where \gls{tddvm} remains close to the failure ratio of Fixed-$100\%$DV with \gls{bla}. Both curves lie lower since \gls{bla} always transmits at maximum power and clears the queued traffic faster.

\subsection{Analysis of Drain Voltage Selection}
\label{subsec:vddsel}

\begin{figure}[!t]
  \vspace{-6pt}
  \centering
  \includegraphics[width=0.93\columnwidth]{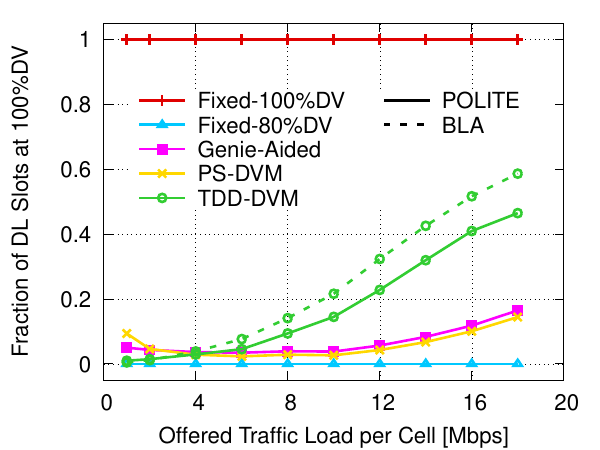}
  \caption{Fraction of \gls{dl} slots at $100\%$DV for each scheme against the offered traffic load per cell.}
  \label{fig:vddsel}
\end{figure}

Fig.~\ref{fig:vddsel} shows the fraction of \gls{dl} slots spent at $100\%$DV against the offered traffic load per cell, which explains how each scheme reaches the consumption of Fig.~\ref{fig:consumed}. The two fixed references mark the extremes of always and never selecting $100\%$DV. The Genie-Aided bound stays at $80\%$DV over most of the range and selects $100\%$DV in only about $16\%$ of the slots at the highest traffic load, with \gls{psdvm} close by. The proposed \gls{tddvm} raises the voltage much sooner and reaches about $47\%$ under \gls{polite} at the highest load. The resource load $\Lambda(t)$ measures the backlog of the whole period and cannot identify the few slots whose output power requires $100\%$DV, so meeting the delivery constraint forces a conservative selection toward $100\%$DV. This surplus of $100\%$DV slots is what separates \gls{tddvm} from the Genie-Aided bound in Fig.~\ref{fig:consumed}. Under \gls{bla}, \gls{tddvm} selects $100\%$DV even more often, since every transmission is served at maximum power and more slots require the full drain voltage. Lowering the drain voltage saves energy only where the transmit power leaves room for it, which \gls{polite} provides and \gls{bla} does not. This is why \gls{polite} enables the gain of \gls{tddvm}.

%% file: sections/5_conclusions.tex
\section{Conclusion}
\label{sec:conclusion}

In this paper, we proposed \gls{tddvm}, a drain voltage modulation scheme that selects the \gls{pa} drain voltage level once per \gls{tdd} period based on the resource load. Operating on this slow timescale keeps the hardware cost of drain voltage switching low, and placing the voltage change in the \gls{ul} period allows a slow ramp without signal distortion. By selecting the lower drain voltage when the resource load permits, \gls{tddvm} improves the \gls{pa} energy efficiency. This is enabled by \gls{polite} \gls{la}, which lowers the transmit power at partial load and moves the \gls{pa} into a regime where a lower drain voltage suffices. At equal delivered throughput, \gls{tddvm} with \gls{polite} reduces the consumed \gls{pa} power by up to $35\%$ with respect to the full drain voltage under \gls{bla}, and by up to $7\%$ with respect to it under \gls{polite}, capturing most of the Genie-Aided gain at moderate loads and roughly half at the highest load, at a fraction of the switching cost and without increasing the packet failure ratio. The remaining gap to the bound stems from the resource load metric, which cannot identify the few slots that require the full drain voltage, and future work will investigate selection metrics closer to the actual power demand.